\documentclass[pra,showpacs,twocolumn,aps]{revtex4-1}

\usepackage{graphicx}
\usepackage{mathrsfs}
\usepackage{bbm}
\usepackage{bbold}
\usepackage{amsfonts,amssymb,amsmath}
\usepackage{slashed}
\usepackage[]{units}
\usepackage{color}
\usepackage[section]{placeins}

\usepackage{subcaption}
\usepackage{natbib}

\newcommand{\eps}{\varepsilon}

\newcommand{\trm}[1]{\textrm{#1}}
\newcommand{\tsf}[1]{\textsf{#1}}

\newcommand{\be}{\begin{equation}}
\newcommand{\ee}{\end{equation}}
\newcommand{\bi}{\begin{itemize}}
\newcommand{\ei}{\end{itemize}}
\newcommand{\bea}{\begin{eqnarray}}
\newcommand{\eea}{\end{eqnarray}}

\newcommand{\sig}{\varsigma}

\newcommand{\w}{\mathrm{w}}

\newcommand{\figref}[1]{Fig. \ref{#1}}

\newcommand{\eqnref}[1]{Eq. (\ref{#1})}

\newcommand{\RNum}[1]{\uppercase\expandafter{\romannumeral #1\relax}}

\def\lambdabar{\protect\@lambdabar}
\def\@lambdabar{%
\relax \bgroup
\def\@tempa{\hbox{\raise.73\ht0
\hbox to0pt{\kern.2\wd0\vrule width.7\wd0
height.1pt depth.1pt\hss}\box0}}%
\mathchoice{\setbox0\hbox{$\displaystyle\lambda$}\@tempa}%
{\setbox0\hbox{$\textstyle\lambda$}\@tempa}%
{\setbox0\hbox{$\scriptstyle\lambda$}\@tempa}%
{\setbox0\hbox{$\scriptscriptstyle\lambda$}\@tempa}%
\egroup }

\definecolor{ashgrey}{rgb}{0.7, 0.75, 0.71}

\newcommand{\ud}{\mathrm{d}}
\newcommand{\LCm}{{\scriptscriptstyle -}} 

\newcommand{\LCperp}{{\scriptscriptstyle \perp}}

\begin{document}

\title{On beam models and their paraxial approximation}

\date{\today}
\author{W.~J.~Waters}
\affiliation{Centre for Mathematical Sciences, Plymouth University, Plymouth, PL4 8AA, United Kingdom}
\email{william.waters@plymouth.ac.uk}
\author{B.~King}
\affiliation{Centre for Mathematical Sciences, Plymouth University, Plymouth, PL4 8AA, United Kingdom}
\email{b.king@plymouth.ac.uk}

\begin{abstract}
We derive focused laser pulse solutions to the electromagnetic wave equation in vacuum. After reproducing beam and pulse expressions for the well-known paraxial Gaussian and axicon cases, we apply the method to analyse a laser beam with Lorentzian transverse momentum distribution. Whilst a paraxial approach has some success close to the focal axis and within a Rayleigh range of the focal spot, we find that it incorrectly predicts the transverse fall-off typical of a Lorentzian. Our vector-potential approach is particularly relevant to calculation of quantum electrodynamical processes in weak laser pulse backgrounds.

\end{abstract}
\maketitle

  \section{Introduction}
As the electromagnetic (EM) field intensities attainable in laser facilities increases, so do the possible applications \cite{daido2012,mourou2012,robson2007} and prospects for studying fundamental physics \cite{dipiazza12,narozhny15}. Example leading high-intensity laser facilities include the VULCAN \cite{patel2005} and HERCULES \cite{yanovsky2008} lasers. Given the variation in laser intensities, applications and configurations, there is an extensive list of different beam and pulse models that describe the electromagnetic (EM) fields produced \cite{mcdonald2000,arlt2000,mazilu2010}. One of the most popular models for describing high-intensity laser beams is the ``Gaussian beam'', and several different approaches have been used to describe the fields (an overview can be found in \cite{Salamin06,mcdonald2000,varga1998,cao2002,salamin2006,salamin2007,salamin2015}). A particularly useful approximation employed to describe on-axis phenomena within the central Rayleigh range of the Gaussian beam, is the so-called ``paraxial approximation''. With the advent of new experimental techniques and the quest for ever higher intensities, focusing of intense laser beams is becoming increasingly important \cite{bahk2005,yanovsky2008}. Higher focusing naturally increases the diffraction angle and brings into question the validity of the paraxial approximation \cite{cao2002,sheppard1999,jeong2015}, especially at sub-wavelength beam waist \cite{salamin2007}.

Other than linearly polarised, a Gaussian beam can also be radially polarised (we refer to this as an ``axicon'' beam). It has been shown that the axicon beam can be focused to tighter spots and has the interesting property of an electric field component in the direction of propagation in the absence of a transverse component on axis \cite{dorn2003,quabis2000,quabis2001}. This longitudinal component of the field gives rise to the applicability of using such a beam in the direct acceleration of particles in the absence of a medium \cite{carbajo2016,dai2011}.

There have been a number of different approaches used to derive the fields of Gaussian beams in the paraxial approximation \cite{mcdonald2000,davis1979,lax1975}. An historical review of paraxial theories is presented by \cite{varga1998}. With the demand for tighter focusing these approaches have been extended, providing more accurate field descriptions using higher order expansions of various small parameters \cite{salamin2007,cao2002,luo2007}. 

There are now over fifty \cite{danson2015,ELI} petawatt laser facilities worldwide as well as even more ambitious facilities planned, with the record intensity being of the order of $10^{22}\textrm{Wcm}^{-2}$ \cite{yanovsky2008}. This combined with sophisticated imaging techniques \cite{koch2003} is providing more experimental evidence of the profile and propagation of high intensity beams. This allows a more accurate theoretical description of high intensity beams and pulses. One such variation observed in some high intensity experiments was that of the transverse intensity profile of the beam spot. It was shown by \cite{patel2005} and further referred to by \cite{yanovsky2008,nakatsutsumi2008} that the profile did not represent that of a Gaussian since only 20\% of the energy was contained within the full width at half maximum of 6$\mu$m and 50\% within 16$\mu$m. Therefore the intensity profile had wide tails which at the intensities used could have an appreciable effect on the target. The suggestion was made by \cite{nakatsutsumi2008} that a Lorentzian or ``\,q-Gaussian\," transverse distribution would better represent the profile observed. It has been considered how such a pulse propagates through plasma and how the beam properties might be affected by a Lorentzian frequency distribution for different q-values \cite{sharma2010}.

Experimental advances have motivated interest in going beyond the ``plane wave model'' of laser-based strong-field quantum electrodynamics (QED) (reviews can be found in \cite{marklund06,dipiazza12,narozhny15,king15a}). On the one hand, this allows testing of the locally-constant field approximation \cite{harvey15} used throughout numerical codes, and on the other hand opens up the possibility of studying new phenomena due to focussing \cite{dipiazza14,dipiazza16,ilderton17a,ilderton17b}, medium effects \cite{becker77,cronstroem77,mendonca11,varro13,raicher15,heinzl16} and non-plane-wave longitudinal structure \cite{king16c}. A focused pulsed laser background can be included in QED calculations perturbatively if it is weak enough, where it enters calculations as the Fourier transform of the vector potential.

The main aim of the current paper is to provide a flexible formulation of the vector potential describing propagating laser pulses. After demonstrating the approach by reproducing well-known results for linearly and radially-polarised Gaussian beams, we apply the method to a model which exhibits non-Gaussian focusing and wider tails in the intensity profile, similar to as in \cite{patel2005}.

This paper is organised as follows. In Sec. \RNum{2} we first outline the method of resolving a vector potential description of focused beams, by reproducing the established results of linearly and radially polarised paraxial Gaussian beams. Following this, we provide a beam description with non-Gaussian transverse profile. In Sec. \RNum{3} we analyse our results for our Lorentzian beam/pulse and compare with the better known Gaussian result. In Sec. \RNum{4} we conclude.

\section{Method}

Our approach is inspired by the works of Coleman \cite{Coleman1982} and Dirac \cite{Dirac1949} and begins with the realisation of the potential $A^{\mu}$ as an ensemble of real photons with momenta $l^{\mu}$. This leads to an expression in terms of a Fourier integral in momentum space:
\be 
\label{APOS}
	A^{\mu}(x) = \int\!\frac{\ud^4 l}{(2\pi)^{4}}~\mbox{e}^{-il\cdot x}\,\widetilde{A}^{\mu}(l) \;,
\ee 
where $A^{\mu}$ is a solution of the wave equation $\square A^{\mu}=0$ and the ansatz $\widetilde{A}^{\mu}(l)$ is dependent on which gauge, polarisation or beam set-up we desire.
\newline

\begin{figure}[!ht]
\noindent\centering
\hspace{0.025\linewidth}
\includegraphics[draft=false,width=0.4\linewidth]{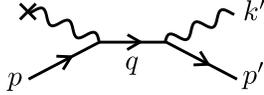}
\caption{Compton scattering in an external electromagnetic field (designated by a cross).}
\label{fig:CS} 
\end{figure}

This is a particularly useful form of the vector potential for calculations in QED when the external field (the laser background) is included perturbatively. Suppose we write the field in terms of the intensity parameter $\xi$ (sometimes referred to as $a_{0}$): $A^{\mu}(x) = (m\xi/e) \epsilon^{\mu} f(x)$ where $|f(x)|\leq 1$, $\epsilon\cdot\epsilon = -1$ and $e>0$ is the charge of the positron. When $\xi\ll 1$, the background field can be included perturbatively. If one considers Compton scattering in such an external field, one of the two leading-order Feynman diagrams is given in \figref{fig:CS}.
Then, we see the scattering matrix element can be written:
\bea
\tsf{S}_{\!fi} = \frac{-i e^{2}}{\sqrt{2^{3}V^{3}p^{0}p'^{0}k'^{0}}}\overline{u}_{p'}\slashed{\eps}_{k'}\frac{\slashed{q}_{\ast}+m}{q_{\ast}^{2}-m^{2}+i\eps}\widetilde{\slashed{A}}(q_{\ast}-p)u_{p},\nonumber \\
\eea
and $q_{\ast} = p'+k'$ i.e. it is the Fourier-transform of the vector potential, evaluated at a momentum determined by in- and out-going particles, which naturally occurs in the calculation.

\subsection{Gaussian Beam}

Starting with \eqnref{APOS} we choose to work in the Lorentz gauge $\partial\cdot A=0$ and set $A^{0}=0$. We begin by considering linear polarisation. For this set-up our ansatz becomes:

\be
\label{aperpansatz}
	\widetilde{A}^\perp(l) = -i \epsilon^\LCperp  E_0(2\pi)^4\, \delta(l\cdot l)\, \rho(l^{0}) \delta_{\epsilon}({l}^{\perp})\, l^3 \theta(l^3 l^0) \;,
\ee 
where $\perp$ represents the combined transverse coordinate. Note that $\widetilde{A}^\perp$ is supported only on the light cone $\delta(l\cdot l)$ with directionality enforced by the $\theta(l^{3}l^{0})$  term, so it is automatically a solution to the electromagnetic wave equation. The $\epsilon^\LCperp$ term is the real polarisation vector, $\rho(l^{0})$ represents the energy spectrum of the photon momenta and in order for the field to be real-valued must obey $\rho(l^{0}) = \rho^{\ast}(-l^0)$. Finally $\delta_{\epsilon}({l}^{\perp})$ is the transverse distribution of the photon frequencies, representing the focusing of the beam. We maintain continuity with the plane wave limit by imposing the condition; 

\bea
\label{PWlimit}
\lim_{\eps\to\infty}\delta_{\eps}(l^{\perp}) = 
\delta(l^{1})\,\delta(l^{2}).
\eea

To recover the Gaussian beam result in the literature we use the ``\,nascent\," delta function of the heat kernel:
\bea
\lim_{\eps\to0}\frac{1}{\sqrt{2\pi\eps}}\,\exp\left[{\frac{-x^{2}}{2\eps}}\right] = \delta(x)\nonumber
\eea 
 to choose:
\bea
\label{deltaregG}
\delta_{\eps}(l^{\perp}) = 
\frac{\w_{0}^{2}}{4\pi}\exp\left[-\frac{\left(l^{\perp}\w_{0}\right)^{2}}{4}\right];
\eea

where $\eps=2/\w_{0}^{2}$ has been chosen and $\w_0$ is the beam waist.
\newline
\smallskip
\newline
To recover the four-potential $A^{\perp}(x)$ and show that this ansatz does indeed reproduce the field of a Gaussian beam as proposed, we must perform the momentum integrals in \eqnref{APOS}. We eliminate the $l^3$ integral using:

\bea
\label{DLL}
	\delta(l\cdot l) &=& \theta\big[(l^{0})^{2}-(l^\perp)^{2}\big]\nonumber\\
	&&\left[\frac{\delta\left(l^{3}-\sqrt{(l^{0})^{2}-(l^\perp)^{2}}\right)}{2|l^{3}|}
+\left(l^{3}\to-l^{3}\right)\right],\nonumber\\
\eea 

where $\theta(\cdot)$ is the Lorentz-Heaviside function.
\newline
\smallskip
\newline
Writing out $A^\perp(x)$ explicitly using \eqnref{DLL} to replace $\delta(l\cdot l)$, we have:

\bea
\label{Aperp}
 A^{\perp}(x) &=& -iE_{0}\frac{\w_{0}^{2}}{8\pi} \int 
dl^{0}\,d^{2}l^{\perp}~
\mbox{e}^{-il^{0}t + 
il^{\perp}\cdot x^{\perp}-\frac{(l^{\perp}\w_{0})^{2}}{4}}\nonumber\\
&&\rho(l^{0})\,\theta\big[(l^{0 
})^{2}-(l^\perp)^{2}\big]\nonumber\\
&& \left[\theta(l^{0})\mbox{e}^{iz\sqrt{(l^{0})^{2}-(l^{\perp})^{2}}}  
-\theta(-l^{0})\mbox{e}^{-iz\sqrt{(l^{0})^{2}-(l^{\perp})^{2}}}\right].\nonumber\\
\eea

 As stated previously, we are looking to connect with the well known analytical result of the Gaussian beam in the paraxial limit. This is achieved using two approximations.
 \newline
 
i) Since we are not solving the paraxial wave equation directly, we must make assumptions about the photon momenta to ensure we recover the same result, namely that:
 
 \be
 \label{paraxassump}
 l^\perp\ll l^0.
 \ee 
 
This allows us to simplify the integral by manipulating the square root term. Performing a Taylor expansion of $l^{3}$ and neglecting terms of order $\left(l^{\perp}/l^{0}\right)^4$ we recover a Gaussian term:

\bea
\label{sqrt}
\sqrt{(l^{0})^{2}-(l^{\perp})^{2}} = |l^{0}|\sqrt{1-\left(\frac{l^{\perp}}{l^{0}}\right)^{2}} 
\approx |l^{0}| - \frac{(l^{\perp})^{2}}{2|l^{0}|}.
\eea

 ii) The integration in $(l^{\perp})^{2}$ is bounded by $(l^{0})^{2}$. Since the Heaviside-Lorentz $\theta$ function only depends on $(l^{\perp})^{2}$, we perform the integral over $l^\perp$ in polar co-ordinates and compute the integral over the angular dependence. Following this we may use the $\theta$ function to determine the limits for the remaining integration:

\bea 
\label{Il0}
I(l^{0},z) = 2\pi \int_{0}^{|l^{0}|} d\rho~\rho ~\mbox{e}^{-a\rho^{2}}\,J_{0}(|x^{\perp}|\rho);
\eea 

where,
\bea 
a = \frac{\w_{0}^{2}}{4}+i\frac{z}{2|l^{0}|}\nonumber
\eea 
and $J_{n}(x)$ is the $n$th order Bessel function of the first kind \cite{olver97}. We find:

\bea
\label{Aperp1}
A^{\perp}(x) &=& -iE_{0}\frac{\w_{0}^{2}}{8\pi} \int 
dl^{0}~
\rho(l^{0})\mbox{e}^{-il^{0}t}\nonumber\\
&&\left[\theta(l^{0})\mbox{e}^{i|l^{0}|z}I(l^{0},z) - \theta(-l^{0})\mbox{e}^{-i|l^{0}|z}I(l^{0},-z)\right].\nonumber\\
\eea

To perform the $l^0$ integral we must state the form of the $\rho(l^0)$ function (frequency spectrum). To recover the Gaussian \emph{beam}, we choose:

\be
\label{RHO}
	\rho(l^{0}) =  \frac{1}{|l^0|}\big[ \delta(l^{0}-\omega)+\delta(l^{0}+\omega) \big],
\ee

where $\omega>0$ is the laser beam frequency.
\newline
\smallskip
\newline
Then we find:
\bea
A^{\perp}(x) = -iE_{0}\frac{\w_{0}^{2}}{8\pi\omega}\mbox{e}^{i\omega(z-t)}I(\omega,z) + \trm{c.c.}.
\eea

Now to deal with $I(\omega,z)$ we make the second assumption that $\omega \w_0 \gg 1$ to perform the integral $I(\omega,z)$ analytically, giving:

\bea 
A^\perp = -\frac{E_{0}}{\omega}\frac{\mbox{e}^{-\frac{|x^{\perp}|^{2
} } { \w^{2} }}}{\sqrt{1+\varsigma^{2}}}\sin\left[\omega x^{\LCm}+\tan^{-1}\varsigma-\frac{|x^{\perp }|^ {2}\varsigma}{\w^{2}}\right],\nonumber\\
\eea

where we have defined:

\bea
\sig(z)&=&\frac{z}{z_r};\nonumber\\
\w(z)&=&\w_0\sqrt{1+\sig^2};\nonumber \\
x^\LCm&=&t-z \nonumber
\eea 

and $z_{r}=\omega^{2}\w_{0}/2$ is the usual Rayleigh length.
The well-documented \cite{varga1998,lax1975,davis1979} result for the electric field of a paraxial Gaussian beam using $E^\LCperp = -\partial_t A^\LCperp$, is then

\be\begin{split}
\label{EPERPG}
	E^{\perp}_{\trm{beam}}(x) &= E_0\frac{e^{-\frac{|x^{\perp}|^2}{\w^2}}}{\sqrt{1+\sig^2}} \cos \bigg[\omega x^\LCm +\tan^{-1}\sig - \frac{|x^{\perp}|^2\sig}{\w^2}\bigg] \;.
\end{split}
\ee

Finally to achieve an expression for a Gaussian paraxial \emph{pulse} simply adapt the frequency spectrum $\rho(l^{0})$ to the desired pulse profile of a Gaussian distribution:

\bea 
\label{RHOP}
\rho(l^{0})=\frac{\tau}{|l^{0}|\sqrt{4\pi }}\left(\exp\left[\frac{-\tau^{2}\left(l^{0}-\omega\right)^{2}}{4}\right]+\left(\omega\to-\omega\right)\right),\nonumber\\
\eea 

Where $\tau$ is the pulse duration. On substituting into \eqnref{Aperp1} and computing, we yield an expression for the Gaussian paraxial pulse which differs from the beam \eqnref{EPERPG} only in the addition of a Gaussian temporal envelope:

\be\begin{split}
\label{EPERPGp}
	E_{\trm{pulse}}^{\perp}(x) &= e^{-\left(\frac{x^\LCm}{\tau}\right)^{2}}E^{\perp}_{\trm{beam}}(x)\;.
\end{split}
\ee

\subsection{Axicon Beam}

To demonstrate the flexibility of this method, we determine the field of an axicon beam. Taking our lead from \cite{mcdonald2000} we consider the case where $A^{3}$ is the only non-zero spatial component of $A$ and, choosing to work in the Lorentz gauge $\partial\cdot A = 0$ implies that $\partial_0 A^{0}=\partial_3 A^{3}$.

The method is the same as before, we evaluate the momenta integrals in \eqnref{APOS} except in this case we have the ansatz:

\be
\widetilde{A}^{3} = -(2\pi)^{4}iE_{0}\delta(l^{2})\rho(l^{0}) 
\delta_{\eps}(l^{\perp})l^{3}\theta(l^{3}l^{0}). \label{a3ansatz}
\ee
 
On performing the integrals we recover the expression for $A^{3}$:

\be
A^3(x) = -\frac{E_{0}}{\omega}\frac{\mbox{e}^{-\frac{|x^{\perp}|^{2
} } { \w^{2} }}}{\sqrt{1+\varsigma^{2}}}\sin\left[\omega x^{-}+\tan^{-1}\varsigma-\frac{|x^{\perp }|^ {2}\varsigma}{\w^{2}}\right].
\ee 

Now to establish the $E^{3}$ component of the electric field we recall that $A^{0}$ is non zero and hence has a contribution to $E^{3}$. Therefore differentiating $A^{3}$ with respect to $z$ gives:

\bea
\partial_{z}A^{3}(x) = 
E_{0}\frac{\mbox{e}^{\frac{-|x^{\perp}|^{2
} } { \w^{2} }}}{\sqrt{1+\varsigma^{2}}}\cos\left[\omega x^{-} + 
\tan^{-1}\varsigma-\frac{|x^{\perp }|^ {2}\varsigma}{\w^{2}}\right].\nonumber\\
\eea

We have neglected a term of order $\left(\omega z_{r}\right)^{-1}$ since $\omega \w_{0}\gg 1$. This implies with the gauge condition:
\bea
A^{0}(x) &=& 
\frac{E_{0}}{\omega}\frac{\mbox{e}^{\frac{-|x^{\perp}|^{2
} } { \w^{2} }}}{\sqrt{1+\varsigma^{2}}}\sin\left[\omega(t-z) + 
\tan^{-1}\varsigma-\frac{|x^{\perp }|^ {2}\varsigma}{\w^{2}}\right].\nonumber \\
\eea 

Therefore $A^{0}(x) = -A^{3}(x)$. 
\newline
\smallskip
\newline

Calculating the electric field using $E^{3} = -\partial_{t}A^{3}-\partial_{z}A^{0}$ and $E^{\perp} = -\partial_{\perp} A^{0} = \partial_{\perp} A^{3}$, and neglecting terms of $O\left((\omega z_r)^{-1}\right)$ we find:

\bea
\label{E3}
E^{3}&=& -\frac{E_{0}}{\sqrt{1+\varsigma^{2}}}\mbox{e}^{\frac{-|x^{\perp}|^{2} } { \w^{2} }}\cos\left[\omega x^{-} + 
\tan^{-1}\varsigma-\frac{|x^{\perp }|^ {2}\varsigma}{\w^{2}}\right],\nonumber\\
\label{Eperp}
E^{\perp}&=&\frac{x^{\perp} E^{3}}{\w^{2}},
\eea 

which agree with recognised \cite{mcdonald2000} expressions for a Gaussian-focused axicon beam. We can see directly from \eqnref{Eperp} that they produce the characteristics associated with an axicon beam: maximum longitudinal \textbf{E}-field component in the centre and the absence of a transverse \textbf{E}-field on-axis. 

\section{Lorentzian Beam}

\subsection{Method}
We seek a beam description that is representative of observations in high intensity experiments showing that the transverse intensity profile is not always well-described by a Gaussian distribution but instead can have wide tails \cite{yanovsky2008,patel2005}. The suggestion by \cite{nakatsutsumi2008} is that a Cauchy/Lorentzian distribution would be more representative. We note that the wide tails can be seen by calculating the average root mean square (RMS) width, which for a Gaussian $\exp\left[-(x/\w)^2\right]$, is $(\sqrt{\langle x^2\rangle}=\w/\sqrt{2})$ whereas for a Lorentzian of the form $1/(1+(x/\w)^2)$ diverges $(\sqrt{\langle x^2\rangle}\to\infty)$.

 We implement a simple variation into our ansatz, demonstrating the flexibility of the method. The function $\delta_{\epsilon}(l^{\perp})$ represents the focusing as it did in the Gaussian case, but we alter its form from Gaussian to a distribution with wider tails. We seek a $\delta_{\eps}(l^{\perp})$ function that produces an intensity profile with wide tails and satisfies the condition (\ref{PWlimit}). We choose the Poisson kernel:

\[
 \lim_{\epsilon\to 0}\frac{2\,\eps}{x^{2}+\eps^{2}} =\lim_{\eps\to 0}\int_{-\infty}^{\infty} \frac{ds}{2\pi}~\mbox{e}^{ixs -|s\,\eps|f[x;\eps]}=\delta(x).
\]

To acquire a beam that is symmetric under rotations about the propagation axis, we set $x^2$ to $l^\perp\cdot l^\perp$, square the Poisson kernel and choose $\eps=1/\w_0$. This leads to:

\bea 
\label{Ldeltareg}
\delta_{\eps}(l^\perp)=\frac{4\w_{0}^2}{\left[1+\left(\w_0 l^\perp\right)^2\right]^2}\;,
\eea 

which, unlike the Gaussian beam case, does not have a simple connection to the plane-wave limit as $\lim_{\eps\to 0}\delta_{\eps}(l^\perp) = \delta[(l^1)^2+(l^2)^2]$.

We use the same polarisation set-up and method used in the Gaussian beam derivation. Starting from \eqnref{aperpansatz} with \eqnref{Ldeltareg}, we proceed in the same manner, finding:

\bea
A^{\perp}(x)&{=}&\LCm2iE_{0}\w_{0}^{2}\int dl^{0}~
\rho(l^{0})\theta(l^{0})\mbox{e}^{-il^{0}t +i|l^{0}|z}I_{L}(l^{0},z)\nonumber\\
&&+\trm{c.c.}, \label{AperpL}
\eea

where:

\bea 
\label{IL}
I_{L}(l^{0},z) = 2\pi \int_{0}^{|l^{0}|} d\rho~\frac{\rho\mbox{e}^{-ib\rho^{2}}}{\left[1+(\w_0\rho)^2\right]^2} ~\,J_{0}(|x^{\perp}|\rho),
\eea 

with $ b=z/2|l^{0}|$.
\newline
\smallskip
\newline
Expanding the Gaussian or Lorentzian in \eqnref{IL} for small argument does not give a satisfactorily convergent expression in regions close to the beam axis. However, we can make use of the Bessel multiplication theorem (\cite{watson22}, page 142):
\bea
\label{BesselMT}
J_{\nu}(\lambda z) = \lambda^{\nu} \sum_{m=0}^{\infty} \frac{(-1)^{m}(\lambda^{2}-1)^{m}(z/2)^{m}}{m!}\,J_{\nu+m}(z).\nonumber\\
\eea
Setting $z=|x^{\perp}|/\w_{0}$ so $\lambda=\w_{0}\rho$:
\bea
J_{0}(|x^{\perp}|\rho) &=& \sum_{m=0}^{\infty} \frac{(-1)^{m}((\w_{0}\rho)^{2}-1)^{m}}{m!}\nonumber\\ && 
~\times\,\left(\frac{|x^{\perp}|}{\w_{0}}\right)^{m}\! J_{m}\left(\frac{|x^{\perp}|}{\w_{0}}\right).\nonumber\\
\eea
A benefit of this expansion is that only the $m=0$ term has a non-zero value at the origin. Therefore, we should expect the lowest orders of this expansion to already quite well approximate the paraxial case. Taking the ``\,paraxial\," condition $|x^{\perp}|\ll\w_{0}$, we acquire; 

\bea
\label{AperpLbeam}
A^{\perp}(x) &=& 8\pi E_{0} J_{0}\left(\frac{|x^{\perp}|}{\w_{0}}\right)\nonumber\\
&&\int dl^{0}~\rho(l^{0})\:\textnormal{Im}\left[e^{-il^{0}x^{-}}F\left(|l^{0}|\w_{0},\frac{z}{|l^{0}|\w_{0}^{2}}\right)\right],\nonumber\\
\eea 

where

\bea 
F(\mu,\nu)=\int_{0}^{\mu} du~\frac{u}{[1+u^{2}]^{2}} ~\mbox{e}^{\frac{-i\nu u^{2}}{2}}.
\eea 
Although the integration can be written in terms of Si and Ci functions, the result is not illuminating. For this \emph{beam} result we once again use delta functions for the photon frequency spectrum $\rho(l^0)$ as in \eqnref{RHO} and similarly for a Lorentzian \emph{pulse} we use a Gaussian distribution \eqnref{RHOP}. The \emph{pulse} calculation is not altogether straightforward and requires careful manipulation to produce an analytic paraxial result. With the benefit of hindsight we find that by making the approximation $F(|l^{0}|,z)\to F(\omega, z)$ we remove the $l^{0}$ dependence and can perform the momentum integrals with ease. This approximation is in excellent agreement with the exact result and is based on the assumption that $|l^{0}|$ is peaked around the central frequency $\omega$. 

To analyse the properties of the beam we calculate the energy density $T^{00}$ from the energy-momentum tensor \cite{jackson1999}:

\bea 
T^{\alpha\beta}=\frac{1}{4\pi}\left[g^{\alpha\mu}F_{\mu\lambda}F^{\lambda\beta}+\frac{1}{4}g^{\alpha\beta}F_{\mu\lambda}F^{\mu\lambda}\right].\nonumber
\eea 

The calculation is simplified by considering a linear polarisation and removing the $A^{3}$ component due its negligible contribution. This leads to:

\bea 
T^{00}=\frac{1}{8\pi}\left[\left(\partial_{0}A^{1}\right)^{2}+\left(\partial_{2}A^{1}\right)^{2}+\left(\partial_{3}A^{1}\right)^{2}\right],
\eea 
which can also be written as $T^{00}=\left(E^{2}+B^{2}\right)/8\pi$, which is the the mod-square of the Poynting vector.

 \begin{figure}[!ht]
\noindent\centering
\hspace{0.025\linewidth}
\includegraphics[draft=false,width=0.95\linewidth]{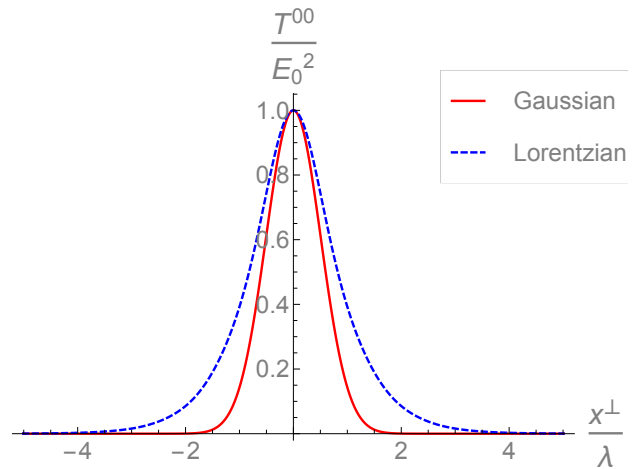}
\caption{Comparison of Gaussian and Lorentzian profile at $\varphi=0$ and $\w_{0}=\lambda$ where the amplitudes have been normalized.}
\label{fig:GLcomp} 
\end{figure}

\subsection{Results}

\begin{figure}[!h]
\noindent\centering
\hspace{0.025\linewidth}
\includegraphics[draft=false,width=0.8\linewidth]{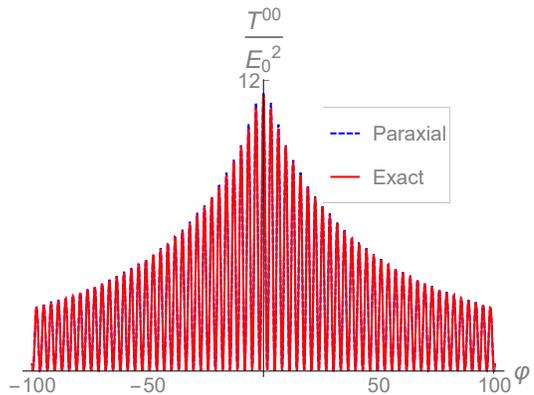}
\caption{Comparison between the paraxial and exact Lorentzian beam solutions on-axis ($x^{\perp}=0$), where $\varphi=\omega x^{-}$.}
\label{fig:onaxiscomp} 
\end{figure}

In this section we analyse the properties of our Lorentzian beam, all plots show the energy density $T^{00}$ with minimum beam waist $\w_{0}=\lambda$ (unless otherwise stated) to accentuate any focusing effects. We first confirm from \figref{fig:GLcomp} that our choice to use a Lorentzian focusing function does indeed produce a beam with wider tails than the Gaussian beam. We further compare with the Gaussian beam by evaluating our paraxial approximation, which we expect to be good on-axis. Since the only dependence on the transverse co-ordinate $x^{\perp}$ was found in the Bessel function \eqnref{AperpLbeam} we find that the ``\,paraxial\," result is very accurate on-axis, ($J_{0}(0)=1$), hence the second paraxial approximation using the Bessel multiplication theorem \eqnref{BesselMT} does not apply. This is shown in \figref{fig:onaxiscomp}.

 \begin{figure}[!h]
 \noindent\centering
\hspace{0.025\linewidth}
\includegraphics[draft=false,width=0.75\linewidth]{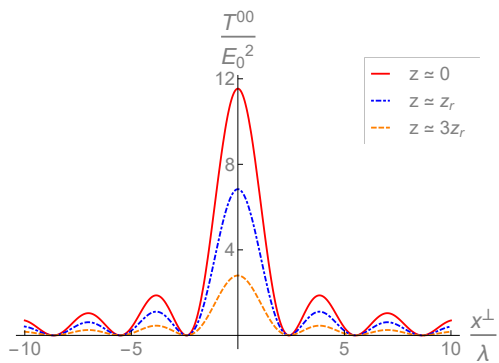}
\caption{Cross section plots of $T^{00}$ for paraxial Lorentzian beam at $t=0$ and with $\w_{0}=\lambda$.}
\label{fig:L2Dparax} 
\end{figure}
\begin{figure}[!h]
\noindent\centering
\hspace{0.025\linewidth}
\includegraphics[draft=false,width=0.75\linewidth]{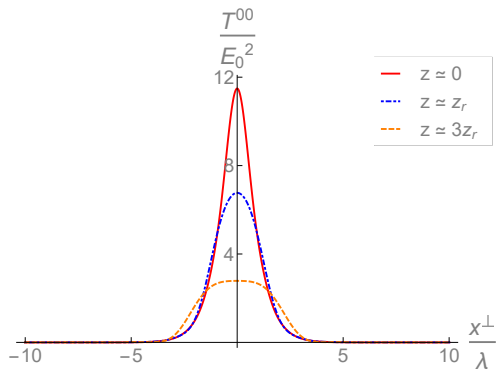}
\caption{Cross section plots of $T^{00}$ for exact Lorentzian beam at $t=0$ and with $\w_{0}=\lambda$.}
\label{fig:L2Dexact} 
\end{figure}

Off-axis however, there are significant variations between the paraxial and exact result. Firstly we observe from \figref{fig:L2Dparax} that the paraxial transverse energy density profile has wide oscillating tails representative of the Bessel function $J_{0}(x^{\perp}/\w_{0})$, whereas the exact beam result \figref{fig:L2Dexact} exhibits a definite width and is non-oscillatory. This is to be expected since the paraxial approximation we made had the effect of removing the transverse momentum dependence from the integrand but gave the paraxial approximation with a $J_{0}(x^{\perp}/\w_{0})$ envelope. 
\begin{figure}[h!!]
\begin{subfigure}[h]{\linewidth}
\includegraphics[width=0.6\linewidth]{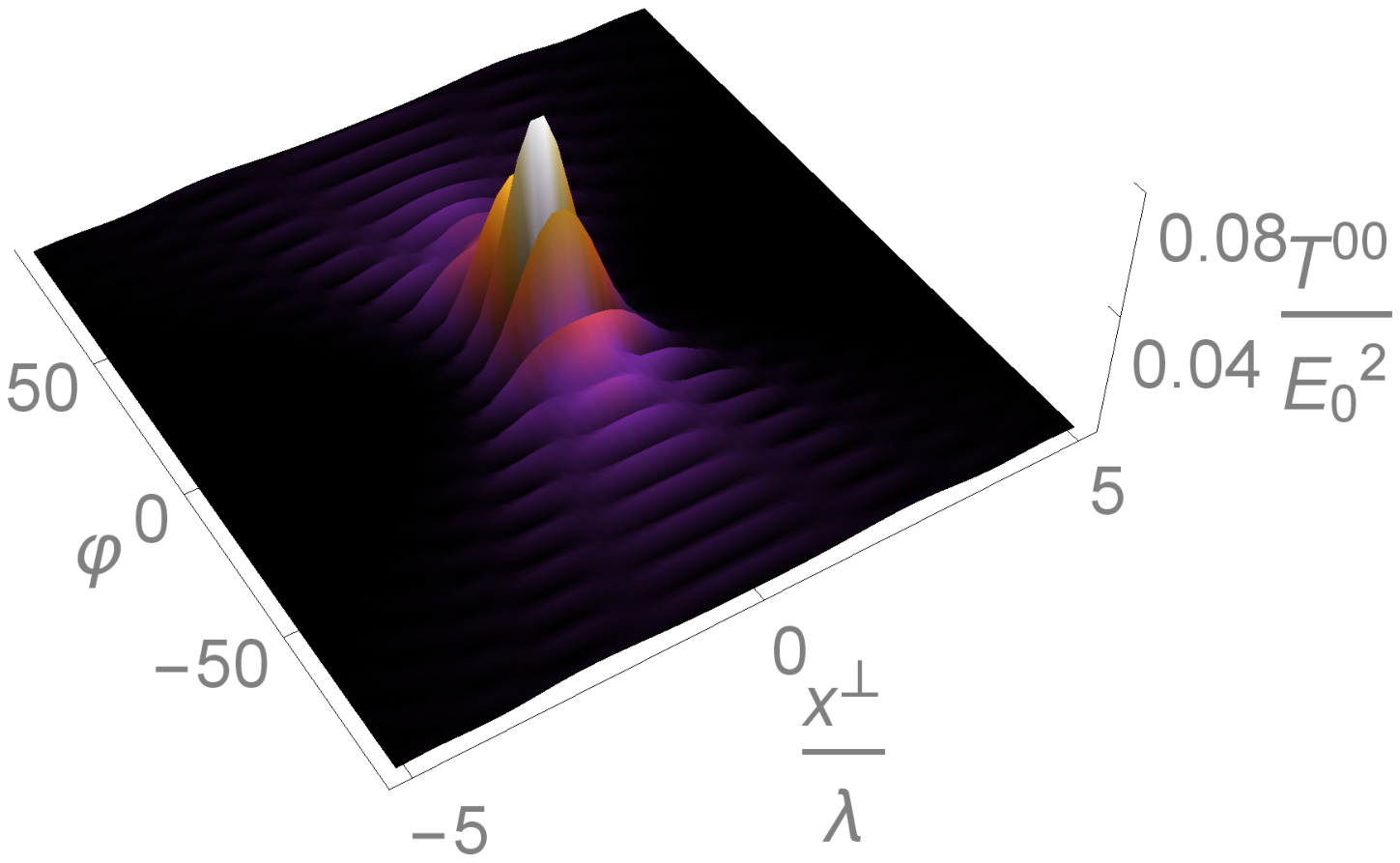}
\caption*{(a)}
\end{subfigure}
\begin{subfigure}[h]{\linewidth}
\includegraphics[width=0.6\linewidth]{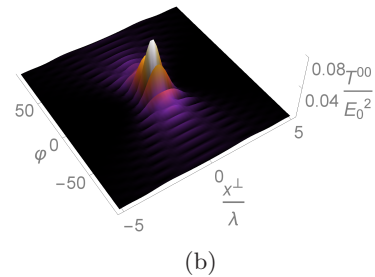} 
\caption*{(b)}
\end{subfigure}
\caption{Profile of (a) paraxial and (b) exact Gaussian beams with $\w_{0}=\lambda$.}
\label{fig:3DGbeam}
\end{figure}

\begin{figure}
\begin{subfigure}[h!!]{\linewidth}
\includegraphics[width=0.5\linewidth]{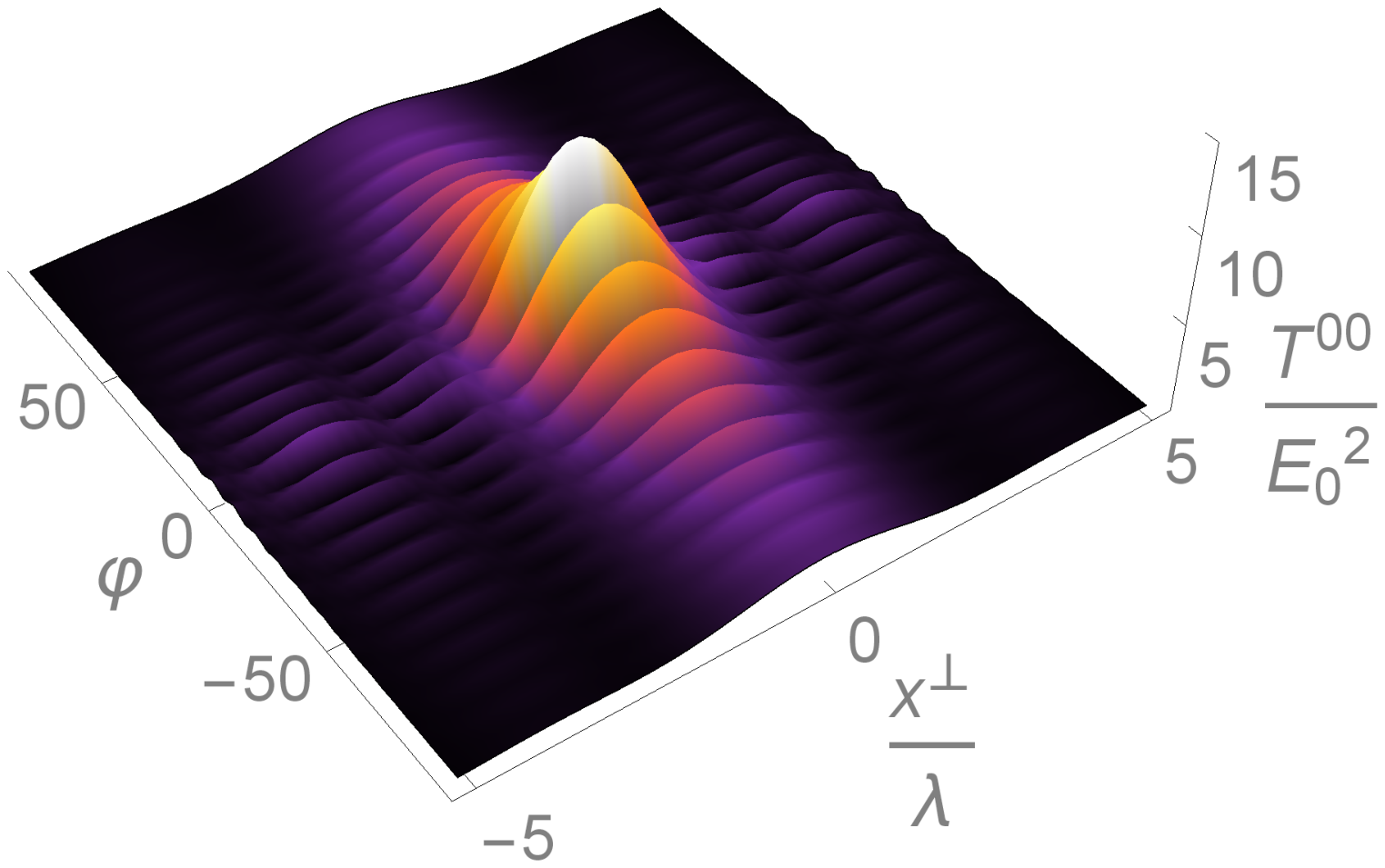}
\caption*{(a)}
\end{subfigure}
\hspace{0.025\textwidth}
\begin{subfigure}[h]{\linewidth}
\includegraphics[width=0.5\linewidth]{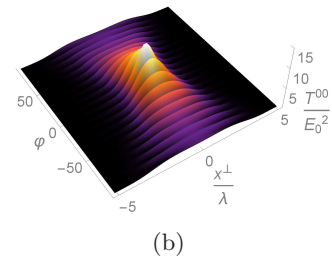}
\caption*{(b)}
\end{subfigure}
\caption{Profile of (a) paraxial and (b) exact Lorentzian beams with $\w_{0}=\lambda$.}
\label{fig:3DLbeam}
\end{figure}

The paraxial approximation presented is leading order, but including higher order terms in the Bessel approximation would have a damping effect on the oscillations off-axis. It is worth noting that the first paraxial approximation \eqnref{sqrt} has little effect for $\w_{0}=\lambda$ as is the case for the Gaussian beam \figref{fig:3DGbeam}. Hence to improve the paraxial Lorentzian result significantly, higher order terms in the Bessel approximation should be included. This Bessel envelope has a significant effect on the beam shape. As observed from \figref{fig:3DLbeam}, the exact solution has the expected shape of a focused beam (a narrowing width towards the focus), however for the paraxial case we observe that $T^{00}$ seems to be bigger than it should be at an appreciable transverse distance from the focus,  near the temporal peak. This has the effect of a beam broadening towards the centre, i.e. the beam will have a convex rather than concave shape. This brings into question the validity of using this leading-order paraxial approximation for describing any off-axis phenomena. 
Therefore, we consider more closely whether the paraxial approximation is representative within a certain regime. In \figref{fig:compz} we plot how the field depends on transverse co-ordinate, in the plane of constant longitudinal co-ordinate, for three different cases. We see that the main peak at the centre of the paraxial profile is a good approximation within one Rayleigh length ($z_{r}$). However due to the absence of width broadening for the paraxial case, we see that the approximation becomes poorer for $z > z_{r}$. \figref{fig:exact3D} displays profiles of the Lorentzian beam for the same beam cross sections. 

\begin{widetext}
\begin{figure*}
\begin{subfigure}[!h]{0.32\textwidth}
\includegraphics[width=0.8\textwidth]{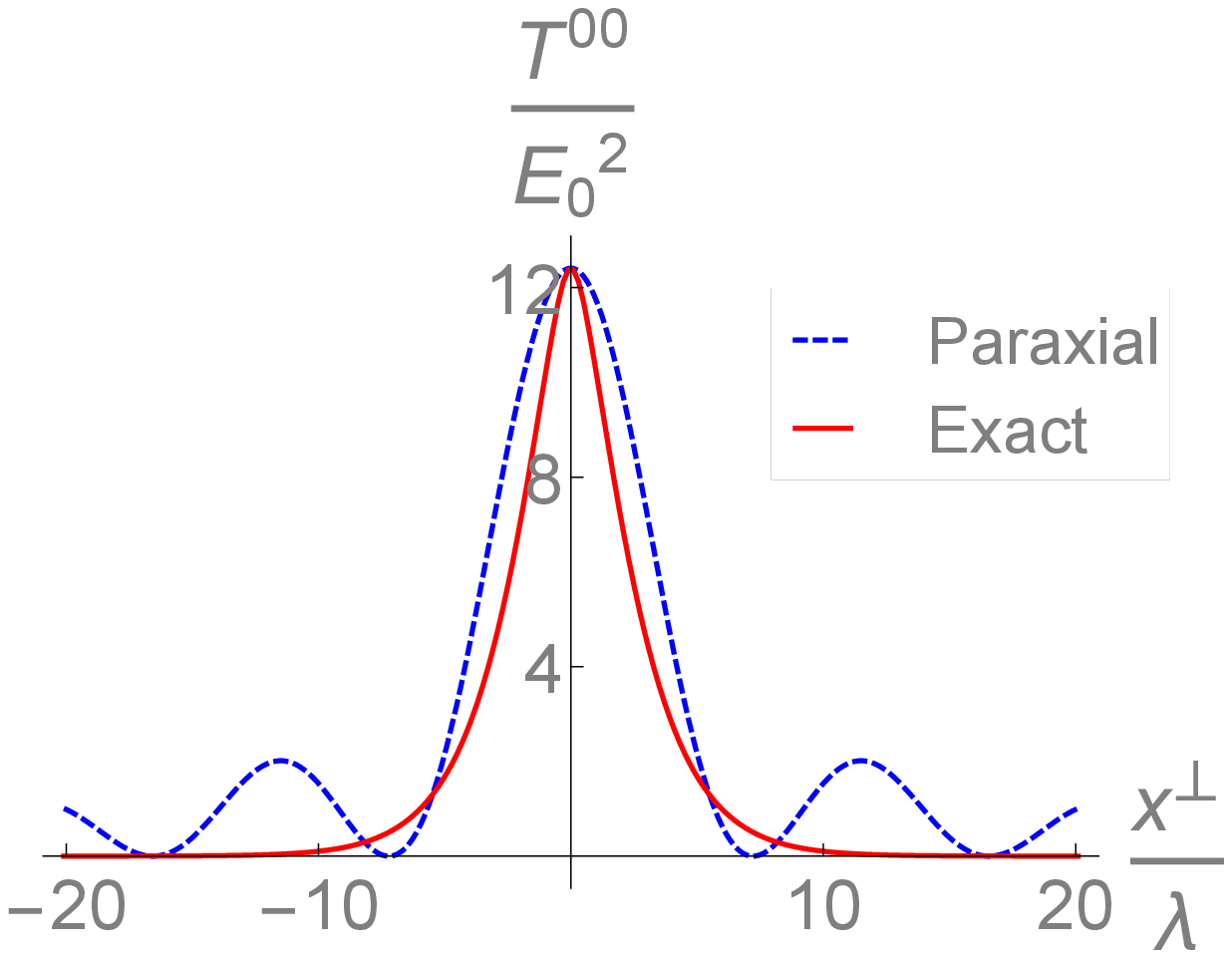}
\caption*{(a)}
\end{subfigure}%
\begin{subfigure}[!h]{0.32\textwidth}
\includegraphics[width=0.8\textwidth]{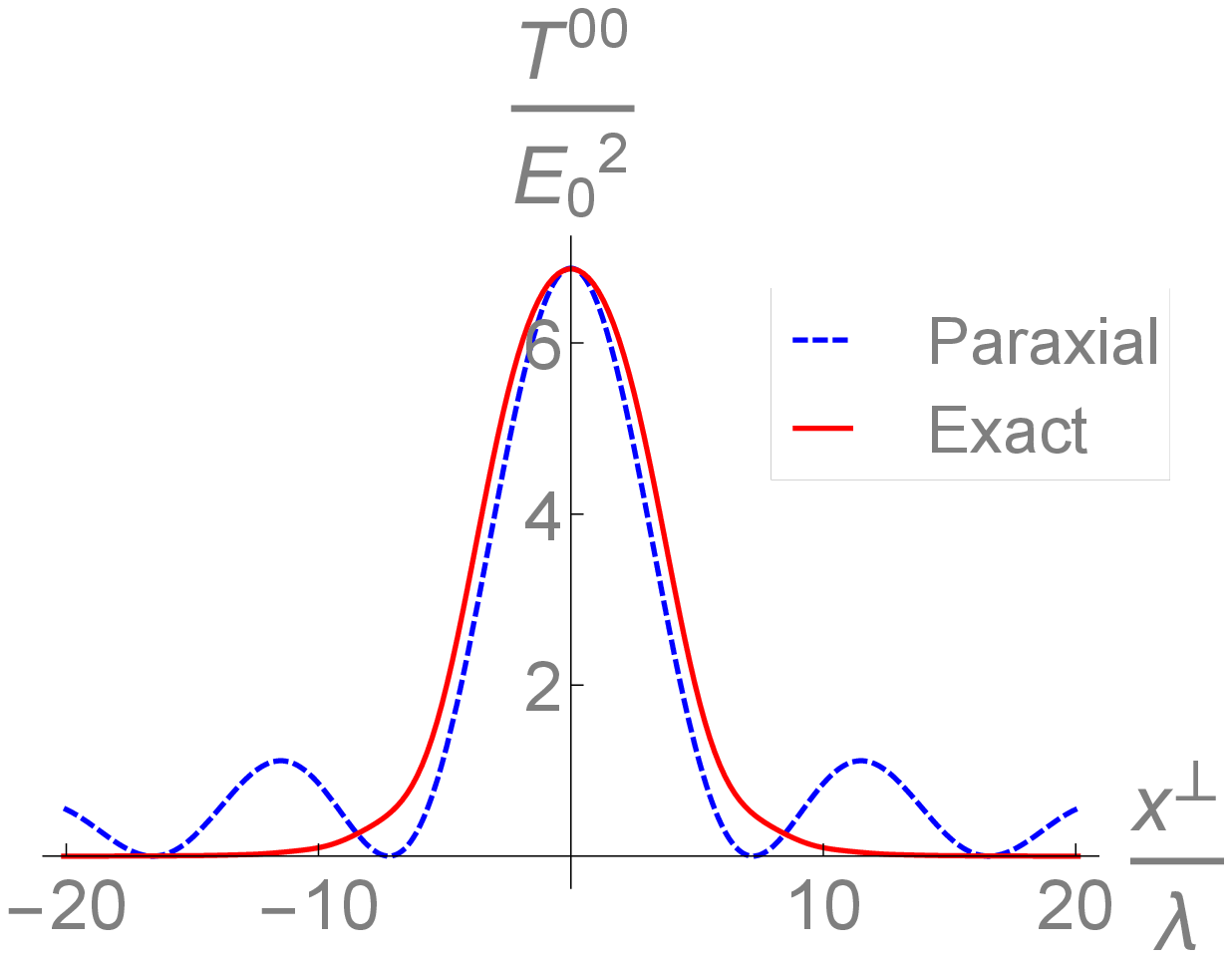}
\caption*{(b)}
\end{subfigure}%
\begin{subfigure}[!h]{0.32\textwidth}
\includegraphics[width=0.8\textwidth]{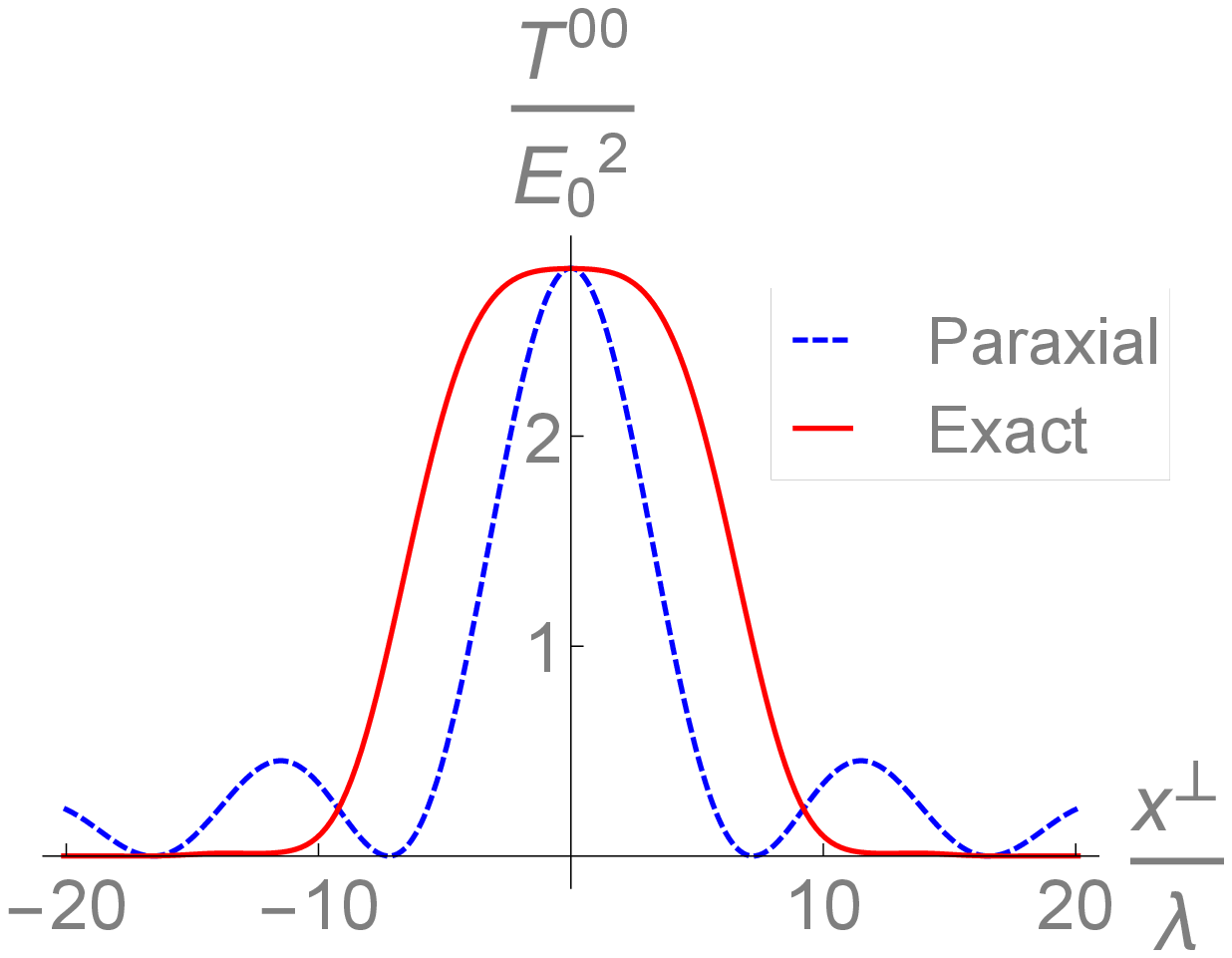}
\caption*{(c)}
\end{subfigure}
\caption{Lorentzian beam cross section comparison at (a) $z= 0$, (b) $z\simeq z_r$ and (c) $z\simeq 3z_r$ for $\w_{0}=3\lambda$ at $t=0$.}
\label{fig:compz}
\end{figure*}

\begin{figure*}
\begin{subfigure}[h]{0.3\textwidth}
\includegraphics[width=\textwidth]{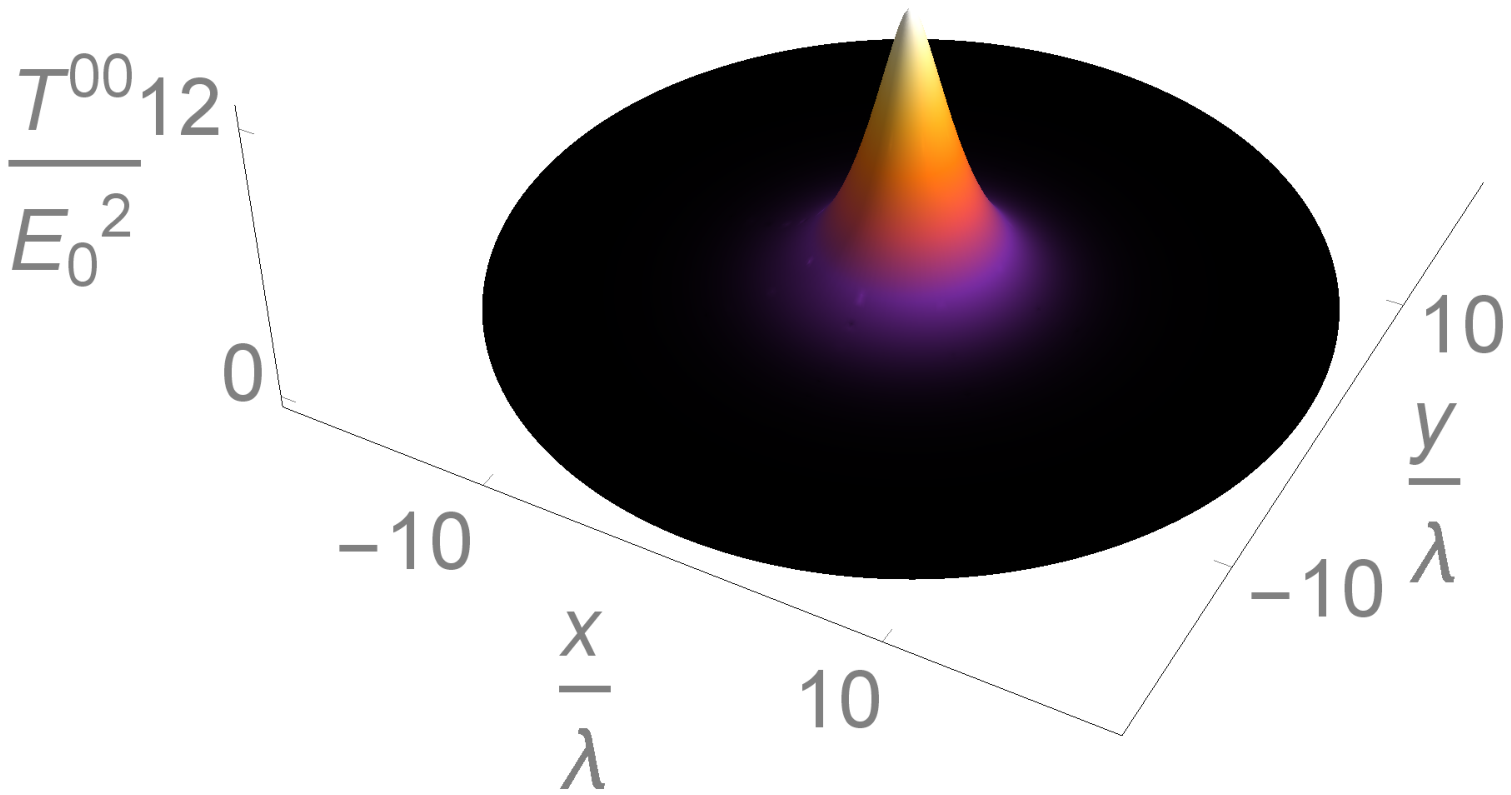}
\caption*{(a)}
\end{subfigure}
\hspace{0.025\textwidth}
\begin{subfigure}[h]{0.3\textwidth}
\includegraphics[width=\textwidth]{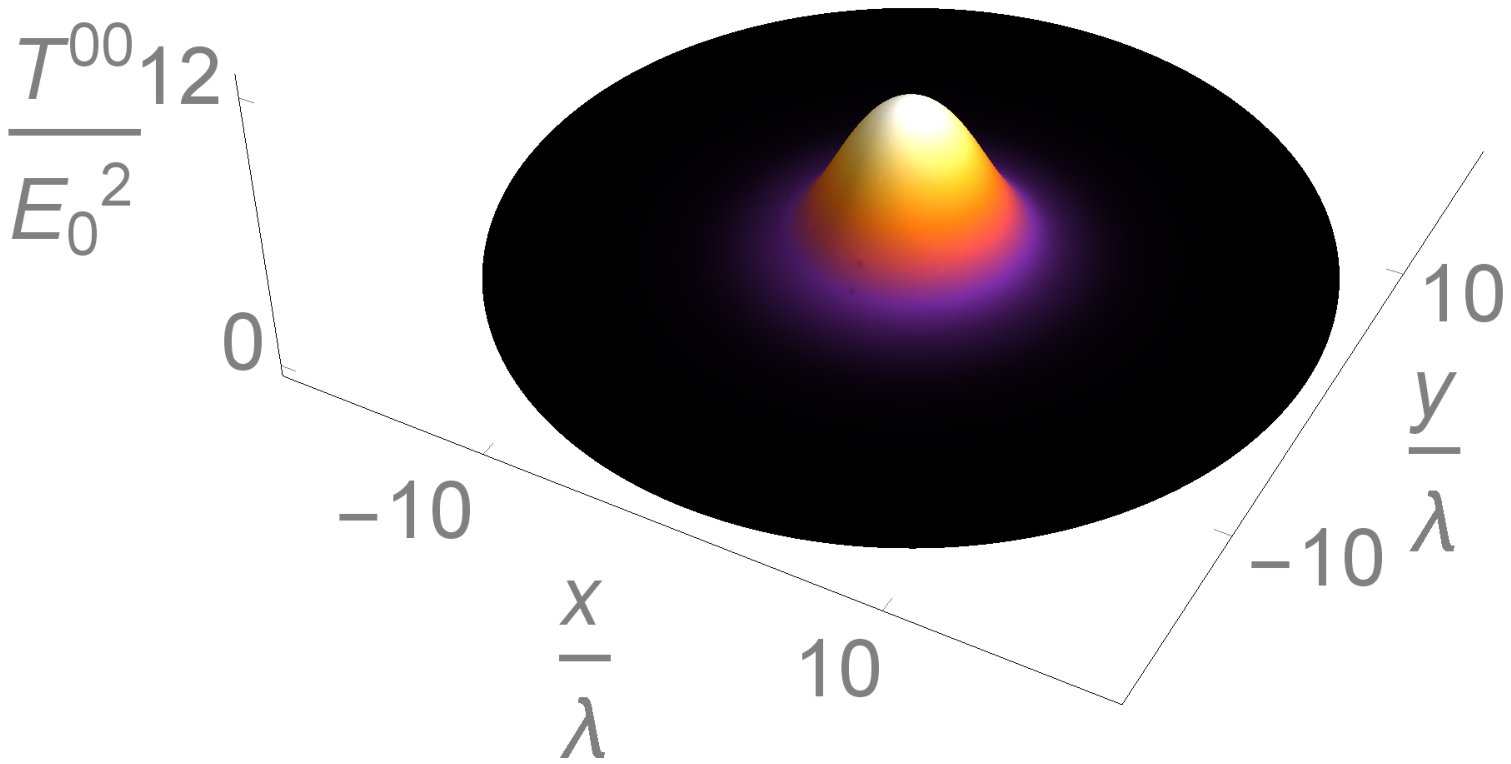}
\caption*{(b)}
\end{subfigure}
\hspace{0.025\textwidth}
\begin{subfigure}[h]{0.3\textwidth}
\includegraphics[width=\textwidth]{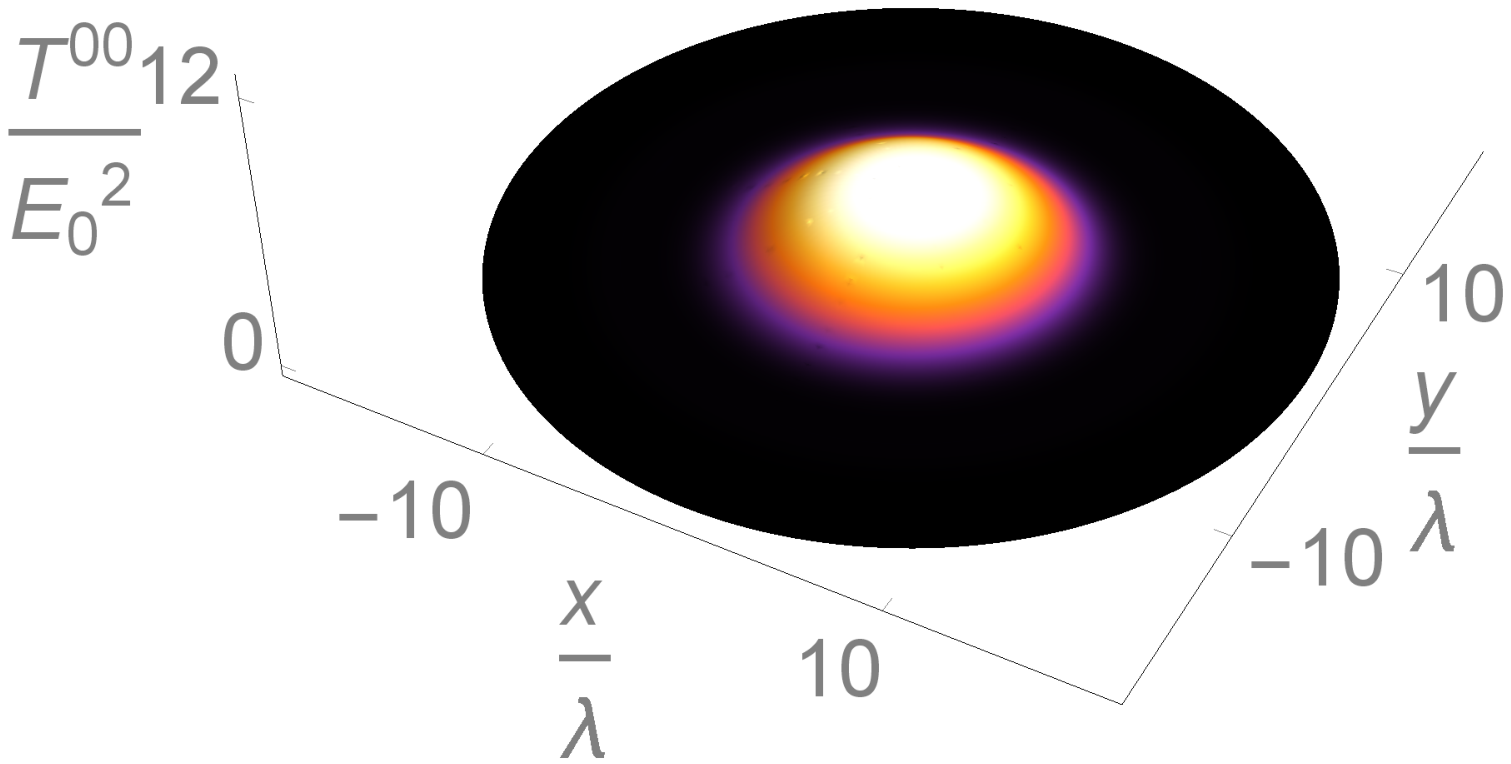}
\caption*{(c)}
\end{subfigure}
\caption{Energy density profile of Lorentzian beam at (a) $z=0$, (b) $z\simeq z_r$ and (c) $z\simeq 3z_r$ for $\w_{0}=3\lambda$ at $t=0$.}
\label{fig:exact3D}
\end{figure*}
\end{widetext}

\section{Conclusion}
We have used a flexible approach based on the Fourier transform of the vector potential in order to derive the fields of focused laser pulses. Using an exact solution to the wave equation, one is able to specify the frequency and transverse momentum distributions to produce the spatio-temporal form of the required focused beam or pulse. Having reproduced the well-known linearly and radially-polarised Gaussian beam results, we applied the method to study a beam with a Lorentzian transverse momentum distribution, as an example of a laser background with wider tails, which is representative of measurements in high intensity laser experiments \cite{patel2005,yanovsky2008,nakatsutsumi2008}. A paraxial approximation was found, which showed excellent agreement with the exact numerical result within the Rayleigh range of the focal spot. The further away from the focal spot, the less accurate the approximation became, and even within the Rayleigh range, oscillating transverse tails were predicted beyond the width of the exact solution.

The Fourier-transformed vector potential formulation of well-known laser pulse backgrounds is particularly useful for QED calculations of laser-particle interactions in the perturbative regime. In particular, the approach demonstrated allows for a flexible and accurate description of high-intensity fields observed in experiment \cite{patel2005,yanovsky2008,nakatsutsumi2008}.

\acknowledgments
The authors would like to thank A. Ilderton for useful discussions, calculations and careful reading of the manuscript. B. K. acknowledges funding from Grant No. EP/P005217/1.

\FloatBarrier

\bibliography{Introbib}
\end{document}